# Nonlinear Elasticity in Biological Gels

Cornelis Storm[1*], Jennifer J. Pastore[2], Fred C. MacKintosh[3], Tom C. Lubensky[1,2] and Paul A. Janmey[1,2]

[1]Department of Physics and Astronomy, University of Pennsylvania

209 S 33rd St, Philadelphia, PA 19104

[2]Institute for Medicine and Engineering, University of Pennsylvania

3340 Smith Walk, Philadelphia, PA 19104

[3]Division of Physics and Astronomy, Vrije Universiteit Amsterdam

De Boelelaan 1081, 1081HV Amsterdam, The Netherlands

---

[*] Present address : Physicochimie Curie, Institut Curie Section de Recherche, 26 Rue d'Ulm 75248 Paris Cedex 05, France.

Unlike most synthetic materials, biological materials often stiffen as they are deformed. This nonlinear elastic response, critical for the physiological function of some tissues, has been documented since at least the 19th century, but the molecular structure and the design principles responsible for it are unknown. Current models for this response require geometrically complex ordered structures unique to each material. In this Article we show that a much simpler molecular theory accounts for strain stiffening in a wide range of molecularly distinct biopolymer gels formed from purified cytoskeletal and extracellular proteins. This theory shows that systems of semi-flexible chains such as filamentous proteins arranged in an open crosslinked meshwork invariably stiffen at low strains without the need for a specific architecture or multiple elements with different intrinsic stiffnesses.

In many network-forming biomolecular systems, the response to a deforming force depends strongly on the magnitude of the deformation, or strain. The process of *strain stiffening*, whereby the material stiffness quantified by its elastic modulus increases with increasing strain, allows biological tissue to respond adaptively to varying external mechanical conditions. Examples of such materials include blood vessels [1], mesentery tissue [2], lung parenchyma [3], and cornea [4]. An additional example that we highlight is the blood clot, composed mainly of fibrin, a filamentous protein that forms branched, crosslinked networks. Strain stiffening allows these networks to be compliant at normal strain levels and strengthen at larger deformations that could threaten tissue integrity. Fibrin is one of several purified biomaterials that display strain stiffening and the structure of the filamentous strands within two such networks composed of neuronal intermediate filaments and fibrin protofibrils are shown in Figure 1. These networks have large solvent-filled spaces, and the filaments are only gently curved between points of intersection. The shear moduli for a variety of biomolecule networks are graphed as a function of applied strain in Fig. 2, and other studies have reported similar strain stiffening in crosslinked actin [5,6], keratin [7], and cytoplasmic gels [5]. Such results are consistent with the finding that internal stresses generated by intracellular motors increase cell stiffness [8].

As Figure 2 illustrates, the systems we consider vary over orders of magnitude not only in their moduli, but also in the strains they tolerate. For example, neurofilament networks can be deformed by well over 400% before failing, while actin networks rupture at strains of only 20%. In this paper, we report a simple theory that can describe these systems, while revealing universal stress-strain relations at low to intermediate strains where entropic forces dominate the elastic response. The input to our theory is the force-extension curve for individual filaments and the assumptions that networks composed of them are homogeneous and isotropic and that their elastic response is affine, that is that the magnitude and direction of local deformations are the same as the overall macroscopic deformation. Our theoretical predictions for the shear modulus as a function of strain agree with the experimental measurements shown in Fig. 2.

**From single semiflexible filaments to networks**

The response of a single elastic filament to an applied force has been the subject of much study over the past years[9-15]. Contrary to simple springs, whose elasticity is primarily enthalpic in origin, the force-response of flexible polymers is dominated by entropy. Because there are many curled-up configurations and only one that is perfectly straight, stretching a flexible filament causes a reduction in conformational entropy and thus an opposing force. The ensuing elastic behavior is inherently nonlinear because for a filament of finite length there is only so much stretching that can be done before one simply runs out of contour. Under most conditions this limit is never approached in networks of rubberlike flexible polymers.

Polymer theory distinguishes three different types of filaments, each characterized by two length scales: the *persistence length* $\ell_p$ (the typical length scale for the decay of tangent-tangent correlations) and the *contour length* $L_c$. A filament is considered *flexible* when $\ell_p \ll L_c$, and *rigid* when the opposite holds. Completely flexible filaments exhibit an elastic response that is completely entropic in origin. Rigid filaments display no entropic elastic response, although *networks* composed of many rigid filaments can and *do* display elastic properties which find their origin in translational and rotational modes of the constituent filaments. Most biologically relevant polymers or filaments, however, are in neither of these classes, but rather in a third intermediate category: that of *semiflexible* filaments with $\ell_p$ and $L_c$ of comparable magnitude. The contour length of these filaments is too small for them to form loops and knots, yet they are sufficiently flexible to have significant thermal bending. In the context of networks, the relevant contour length is that between network junction points or crosslinks, and we shall call the average distance along the filaments between crosslinks in a network $L_c$.

A theory for the force-response of such semiflexible filaments and the corresponding network shear modulus was proposed by MacKintosh *et al*[12]. While they explicitly calculated the modulus only in the linear regime, their model explains qualitatively the non-linear strain stiffening and predicts that the range of strain over which the material has linear elasticity decreases with increasing concentration. This model is based on an energy functional of the form

$$E = \int_0^{L_c} dz \left\{ \frac{\kappa}{2} |\vec{u}''|^2 + \frac{f}{2} |\vec{u}'|^2 \right\}, \tag{1}$$

where $z$ is the projected coordinate along the end-to-end vector, $\kappa$ is the bending stiffness (related to the persistence length as $\kappa = k_B T \ell_p$) and $f$ is the applied force. As shown in Fig. 3, $\vec{u}(z)$ describes the deviation of the filament from its straight conformation. Using the equipartition theorem [16] and the geometric relation between contour length and $\vec{u}(z)$, it is straightforward to calculate the length $L=L(f;L_c)$ of the end-to-end vector to harmonic order in $\vec{u}(z)$. Strictly speaking, this result only holds in the limit of short, stiff filaments or filament segments between crosslinks in a network. The result is most conveniently expressed in terms of the scaled difference between the extension at force $f$ and that at force zero

$$\delta\widetilde{L} = \frac{L(f;L_c) - L(0;L_c)}{L_c^2/\ell_p} = \frac{1}{\pi^2}\sum_{n=1}^{\infty}\frac{\varphi}{n^2(n^2+\varphi)}, \quad (2)$$

where $\varphi = fL_c^2/\kappa\pi^2$ is a dimensionless force (see the inset to Fig. 4). Note that the scaling force $\kappa\pi^2/L_c^2$ is exactly the threshold force for the mechanical Euler (buckling) instability in thin cylinders[17]. Equation (2) can be inverted to yield a force-extension relation that, like the worm-like chain model[9-11], diverges as $f \sim (L - L_c)^2$ as $L \to L_c$.

To get from the force-extension curve of single filaments to the bulk elastic properties of a network of filaments requires a model for the geometry of the network. We consider a model isotropic network, in which pairs of crosslink points (nodes) are connected by links composed of independent semi-flexible filaments. We assume that no torques are exerted at nodes so that filaments stretch or compress but do not bend in response to external forces. We also assume that the network is sufficiently random that the probability distribution for the direction of end-to-end vectors of filaments is isotropic in the absence of external shear stresses. We allow for the possibility of a distribution of end-to-end separations between nodes, with an average equal to $L(f=0;L_c)$ (where $L(f=0;L_c) = L_c(1 - L_c/6\ell_p)$ is the equilibrium end-to-end length of a polymer segment with a total contour length $L_c$) In response to external stresses, network nodes will displace, and filaments will stretch or compress, i.e., the network will strain. Under strain, the position $\mathbf{R}_{0\alpha}$ of node $\alpha$ in the unstrained network will transform to a new position $\mathbf{R}_\alpha = \underset{\sim}{\Lambda}_\alpha \mathbf{R}_{0\alpha}$ where $\underset{\sim}{\Lambda}_\alpha$ is the local (not necessarily symmetric) Cauchy deformation tensor. In random microscopically inhomogeneous yet macroscopically homogeneous networks such as those we consider, $\underset{\sim}{\Lambda}_\alpha$ will in general depend on α, i.e., deformations will be non-affine. We make the enormously simplifying assumption that deformations are affine, i.e., that $\underset{\sim}{\Lambda}_\alpha = \underset{\sim}{\Lambda}$ is independent of α. Though there is no *a priori* reason to believe that this affine approximation is a good one, two recent numerical studies [18,19] of two-dimensional networks of randomly crosslinked semi-flexible rods provide strong evidence that it is, at least at high density and molecular weight. Under affine distortions, the separation **r** of a pair of crosslink points will transform to $\underset{\sim}{\Lambda}\mathbf{r}$, and the end-to-end length of the filament connecting the points will transform from $r = |\mathbf{r}|$ to $|\underset{\sim}{\Lambda}\mathbf{r}|$. Since all forces are transmitted by links, and the force on each link is determined entirely by its length, we can calculate the non-linear stress tensor by averaging the force per area exerted by a single link over all magnitudes and directions of **r**. The result is

$$\sigma_{ij}^T = \frac{\rho}{\det \underset{\sim}{\Lambda}} \left\langle f(|\underset{\sim}{\Lambda}\mathbf{r}|) \frac{(\Lambda_{il} r_l)(\Lambda_{jk} r_k)}{|\underset{\sim}{\Lambda}\mathbf{r}|} \right\rangle_{P(\mathbf{r})},$$

where $\rho$ is the number of links per unit volume and the average is over the probability distribution $P(\mathbf{r})$ for separations **r**. This stress tensor measures the force per unit area of the deformed sample, and, unlike the engineering stress, which measures the force per unit area of the undeformed sample, is symmetric.

We will consider only the volume-conserving simple shear deformation produced in standard rheometers characterized by

$$\underset{\sim}{\Lambda} = \begin{pmatrix} 1 & 0 & \gamma \\ 0 & 1 & 0 \\ 0 & 0 & 1 \end{pmatrix}$$

In this case, the shear modulus is related to the *xz*-component of the stress tensor via $G(\gamma) = \sigma_{xz}(\gamma)/\gamma$.

To evaluate the stress tensor and associated shear modulus, we need to specify the bond length probability distribution and force law *f(L)*. The simplest model is one in which *f(L)* is determined by Eq. (2) and the distribution function is one that sets the distance between all nodes equal to the zero-force end-to-end length $L(f=0; L_c)$ of the link filaments, all of which we assume have the same contour length $L_c$. In this model, no new lengths are introduced by the averaging process defining the stress tensor, and the curve of shear modulus versus strain will exhibit a universal scaling similar to that of the force-extension curve onto which data described by it will collapse. Figure 3 plots experimentally measured shear moduli *G* scaled by their value $G_0$ at zero strain versus strain $\gamma$ normalized by the strain $\gamma_8$ at which *G=8G(0)*. The value of 8 is arbitrarily chosen as the degree of stiffening that all systems examined exhibit before failure, but otherwise has no special significance. There is decent collapse of all data at least at modest strains onto the universal curve predicted by the simple model, suggesting that the initial stages of strain stiffening in biopolymer networks are dominated by entropic effects that are well captured by the theory of semi-flexible polymers.

**Limits of the theory and beyond the entropic regime.**

The observed universality has two important limitations. First, it concerns only the low strain regime where entropic elasticity provides the dominant contribution. While strains of 25% or more appear considerable, the geometry of shear deformation is such that the maximal stretch any filament experiences is only 13%, still well within the entropic regime in most cases. Also evident from Fig 4 is that at high strains departures from the theory become apparent, especially for the vimentin network. Evidently, entropy alone does not capture all the physics of these systems.

We now introduce various modifications of the simple model to improve agreement with experimental data. We first address high-strain deviations from the simple entropic model. We believe these arise from enthalpic contributions to $f(L)$, which we describe via the introduction of a stretch modulus $K$, which measures the longitudinal compliance of our previously inextensible filaments. The force-extension relation Eq. (2) now corresponds to the infinite-$K$ limit of the relation for an extensible semi-flexible filament, which can be expressed in terms of the $K = \infty$ expression as

$$L_K(f; L_c) = \left(1 + \frac{f}{K}\right) L\left(f\left(1 + \frac{f}{K}\right); L_c\right) ,$$

where $L(f; L_c)$ is defined by Eq. (2). The introduction of a stretch modulus is motivated in part by the fact that any continuous object with a finite diameter that can be bent must necessarily be stretchable as well. The resistance of a rod to bending deformations originates in the resistance to stretching and compression produced in it by bend. Therefore the persistence length and the stretch modulus of a rod are not independent, and for a cylindrical tube of radius $r$ they are related through [20] $\ell_p / K = r^2 / 4k_B T$.

The restriction that all filaments have the same end-to-end length even if they have the same contour length is not realistic. Though the actual distribution of lengths depends on how the network is formed, one can expect that crosslinking, however it occurs, creates local pair-wise node separations that differ from the zero-force end-to-end lengths of the filaments. We introduce a specific model for the distribution of end-to-end lengths obtained by generalizing the classical theory of elasticity of crosslinked flexible polymers. We take the distribution function of end-to-end lengths of filaments in our network to be the equilibrium distribution of end-to-end lengths of semiflexible inextensible filaments of contour length $L_c$ [17]

$$P(r) = \frac{2\ell_p}{L_c^2} \sum_{n=1}^{\infty} (\pi n)^2 (-1)^{n+1} \exp\left[-\frac{\ell_p (\pi n)^2}{L_c}\left(1 - \frac{r}{L_c}\right)\right].$$

Since stresses are assumed to be small in the initial configuration, the neglect of stretch elasticity in this distribution is justified. The end-to-end separation between network nodes $L(f=0; L_c)$ should now coincide with the expectation value of the length. An important consequence of having distributed lengths in this manner is that the filaments that make up the network no longer individually experience zero force (as they do for a distribution with all lengths equal to the mean). In the present situation, some will be stretched slightly while others will be compressed relative to their zero-force length $L(f=0; L_c)$. As a consequence, bulk mechanical equilibrium at zero isotropic stress will occur not at zero strain but at an isotropic strain described by an isotropic deformation tensor $\Lambda_{0,ij} = \Lambda_0 \delta_{ij}$. The deformation $\Lambda_0$ is determined by the condition that the isotropic part of the stress tensor $\sigma^I \delta_{ij}$ vanish. The angular average in the expression for the stress tensor in the presence of isotropic deformations is easily calculated to yield

$$\sigma^{I}(\Lambda_0) = \frac{\rho}{3\Lambda_0^2} \int_0^{L_c} dr\, r\, P(r) f(\Lambda_0 L(f=0; L_c)) = 0$$

as the condition determining $\Lambda_0$. This equation is solved numerically for particular parameter values. Note that $\Lambda_0$ is *not* an extra parameter – it is completely determined by the force-extension relation and the distribution function, both of which in turn are functions of the persistence length, the contour length and the stretch modulus only. The condition of zero stress determines $\Lambda_0$.

This extended theory is fit to a series of modulus-strain curves obtained for fibrin at different mass concentrations. Fig. 5 shows the experimental data and the theoretical curves, as well as the best-fit parameter values. Not all parameters were actually fit: the persistence length was determined to be 0.5 μm by calculating the typical length scale for the decay of tangent-tangent correlations based on digitized microscopic images of fibrin filaments such as those in Figure 1. The mesh size $\xi$ of an isotropic network of polymers is $\xi=3\lambda/c$, where $\lambda$ is the mass per unit length of polymer and $c$ is the mass density. Fibrin is composed of fibrin protofibrils in which fibrin monomers of molecular mass 330,000 Da[2], and length 45 nm assemble in a half-staggered arrangement to form filaments of diameter 10 nm[21] with a linear mass density of $\lambda = 2.43*10^{-14}$ mg/μm$^2$. The number of links per unit volume is $\rho=\lambda/cL_c=3/(\xi^2 L_c)$. We assume that $\xi \sim L_c$ so that $\rho=3/\xi^3$. Since $\rho$, $\xi$, and $\Lambda_0$ are all determined by the mass concentration, our model has only one free parameter, the stretch modulus $K$. The best-fit values for K from Fig. 5 are of order 50 to

---

[2] Note that we quote fibrinogen concentrations. Fibrinogen (M=340 kDa) is converted into fibrin before polymerization, which introduces an extra factor of 3.3/3.4 when converting concentrations to monomer masses.

100 pN, very close the elastic-rod estimate of $K= 4k_B T\ell_p/r^2$ for a rod of radius $r=10$ nm and $\ell_p=0.5$ μm.

For all parameter values considered $\Lambda_0$ is smaller than, but of order one, implying a small isotropic compression. This result is a direct consequence of the asymmetric character of the single filament force-extension relation around zero extension (see the inset to Fig. 4). Since the force rises steeply for extensions but remains relatively flat for compressions, stretched filaments contribute more to the residual stress after crosslinking, and the network as a whole will shrink.

The model we have presented makes broad simplifications concerning the role of network geometry. The assumption that crosslinking length and mesh size are equal in particular is needlessly restrictive, and we have also fitted the fibrin curves to a model that includes a separate crosslinking length $\ell_x$, independent of the mesh size. The value of both parameters can in principle be extracted, without loss of generality, by using the initial modulus $G'(0)$ and the shape of the curve at low strain as separate features to fit. Unfortunately, the amount of experimental data points available at low strain is insufficient to draw reliable conclusions from such a fit procedure, and we therefore choose to present only the results of the less general model here.

**Discussion.**

We have shown that nonlinear elasticity, and specifically strain stiffening, is generic to any network composed of semiflexible filamentous proteins. The degree of strain stiffening can be quite large, with 10-fold or larger increases in shear moduli under modest strains as small as 20%. The strain at which stiffening becomes significant depends strongly on the persistence length of the filament and weakly on the mesh size of the network. The stiffest filaments like F-actin or collagen stiffen at a few % strain whereas more flexible filaments like vimentin stiffen only at larger strains, approaching 100%. Unlike the functional form of strain-stiffening at intermediate strain, which is generic for all semiflexible systems, the maximal degree of strain stiffening depends on the molecular details of each polymer since stiffening decreases as the force due to entropic effects becomes comparable to enthalpic effects due to filament extension. Likewise, the strain at which the networks rupture also depends on the nature of the chemical bonds holding the filament together. So for example, F-actin which stiffens at the smallest strain, also ruptures first, usually with a modulus only modestly higher than its low strain limit, because the longitudinal bonds between monomer subunits consist mainly of hydrophobic surfaces orthogonal to the filament axis and are fragile to elongation, whereas fibrin and intermediate filament subunits overlap extensively in staggered arrangements with greater resistance to longitudinal forces.

These results have several implications for cytoskeletal and other biopolymer networks. They suggest that design of artificial biomaterials to replace tissues such as the arterial wall whose function requires non-linear elastic response [1] can be facilitated by polymers of stiffness and mesh size appropriate to produce strain-stiffening at the required range of deformation. As a result of this non-linear passive elasticity, biological systems, like the cytoskeleton can actively manipulate their stiffness by local contraction of the network using motor proteins. For example modest strains exerted by myosin minifilaments in a composite network composed of crosslinked F-actin and intermediate filaments could reversibly increase the stiffness of the cell by a factor of 10 without any change in the degree of polymerization or crosslinking. Once such a network was stiffened, the modulus could rapidly be reduced either by dissociating the stable crosslinks or by relaxing the motors, whose activity may also fluidize these uncrosslinked solutions of cytoskeletal filaments [22].

The inherent strain-stiffening of semiflexible polymer networks is probably only a starting point for the designs that have evolved to endow biological materials with their mechanical properties. Many strain-stiffening tissues are well documented to have strikingly ordered stiff filaments often co-assembled with an amorphous matrix. Variation of such ordered elements in harder material like bone or wood clearly provide other mechanisms to create nonlinear elasticity. The significance of the present work is that such geometrical ordering is not required to produce the high degree of strain stiffening that is an essential aspect of the elasticity of isotropic networks of crosslinked semiflexible polymers.

**Acknowledgements**

We are grateful to Jean-Francois Leterrier, Soren Hvidt, Peter Traub, John Hartwig, and Evelyn Sawyer for collaboration in producing the protein networks and electron mcirographs. Some of this work was initiated by support (of FCM and PAJ) at the Kavli Institute for Theoretical Physics at the University of California, Santa Barbara.

Correspondence and Requests for materials should be addressed to: cornelis.storm@curie.fr

**Figure Legends**

Fig.1: Metal-shadowed neurofilaments and uranyl acetate-stained fibrin protofibrils prepared as described in [23] and [24] respectively, imaged by transmission electron microscopy showing the finite excess of filament contour length between crosslinks and overlap points.

Fig. 2: Dynamic shear storage moduli measured at different maximal strain amplitudes are shown for a series of different crosslinked biopolymer networks. F-actin/filamin gels were prepared as reported in [25]; vimentin (2 mg/ml) was a gift of Dr. Peter Traub and prepared as described in [26]; neurofilaments (3 mg/ml) were prepared as described in [23], rat tail collagen(2 mg/ml) was obtained from Sigma; polyacrylamide/bisacrylamide (5%) was polymerized with ammonium persulfate and TEMED by standard methods; fibrinogen was purified from salmon blood plasma as described in [27] and dialyzed into 50 mM Tris; 450 mM NaCl, pH 8.5, a condition where only fibrin protofibrils form and lateral association of protofibrils into thicker bundles is prevented.

Fig. 3: Definitions and conventions for a single elastic filament. The polymer segment is subject to a force of magnitude $f$ that is applied along the z-direction defined by the end-to-end vector. The field $\vec{u}(z)$ describes the deviation from a straight configuration.

Fig. 4: Data of Fig. 1. scaled as described in the text (symbols) and theory (solid line). The theoretical curve was obtained assuming a uniform distribution of filament lengths

(i.e. all equal to the mesh size), and upon averaging over all orientations in 3 dimensions. Inset: the dimensionless force vs. extension curve described by Eq. (2).

Fig. 5: Experimental data for fibrin protofilaments (dots) at various concentrations (lowest concentration being the lowest-lying curve), and corresponding theoretical curves (solid lines) as computed from the extended theory including a stretch modulus. Best-fit values were determined as

a) c=0.5 mg/ml ($\xi$=0.39 micron): K=67pN, $\Lambda_0$=0.948,

b) c=1.0 mg/ml ($\xi$=0.27 micron): K=58pN, $\Lambda_0$=0.969,

c) c=2.0 mg/ml ($\xi$=0.19 micron): K=73pN, $\Lambda_0$=0.981,

d) c=4.5 mg/ml ($\xi$=0.12 micron): K=110pN, $\Lambda_0$=0.991

**Figures**

Fig. 1:

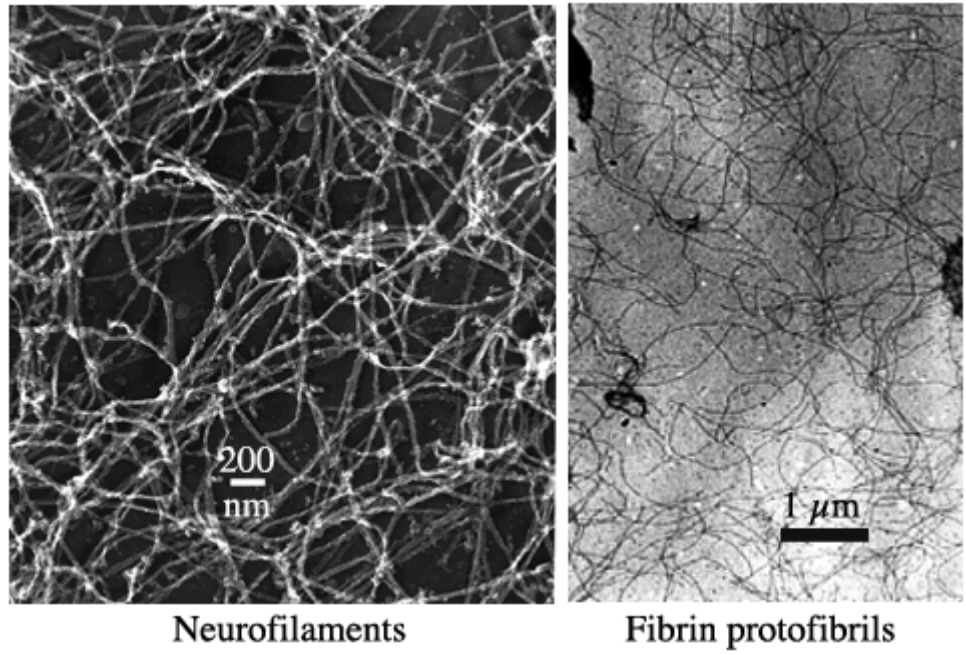

Fig. 2:

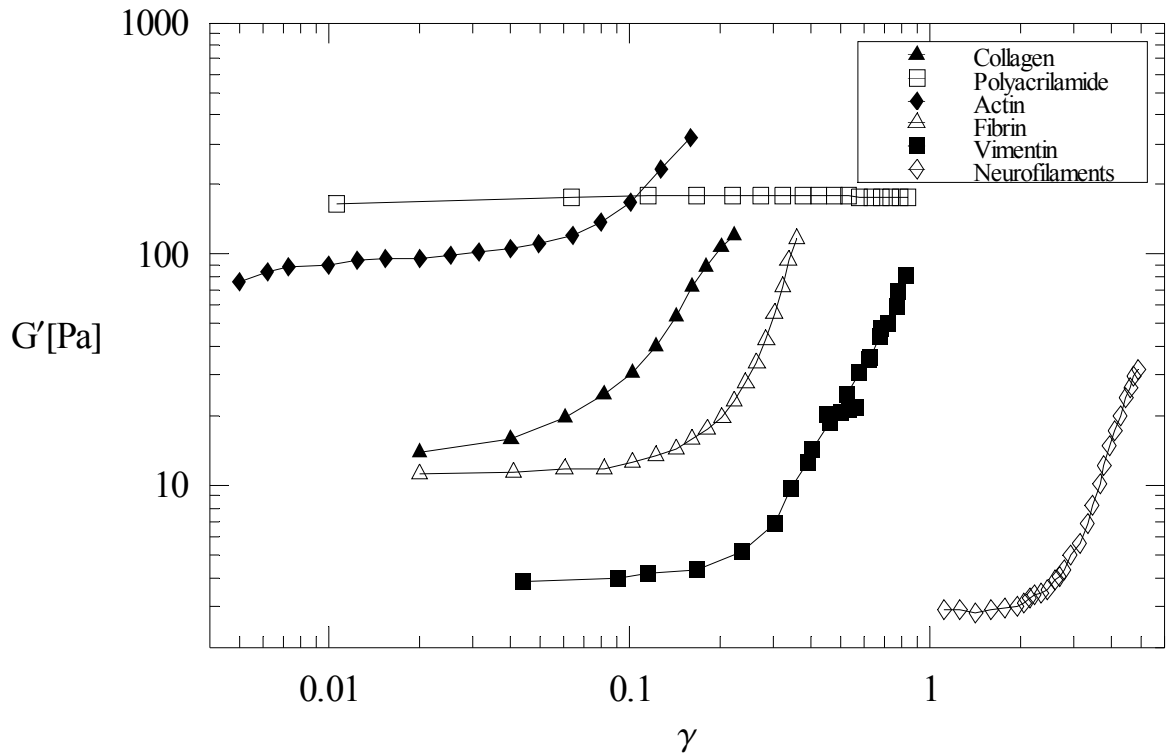

Fig. 3:

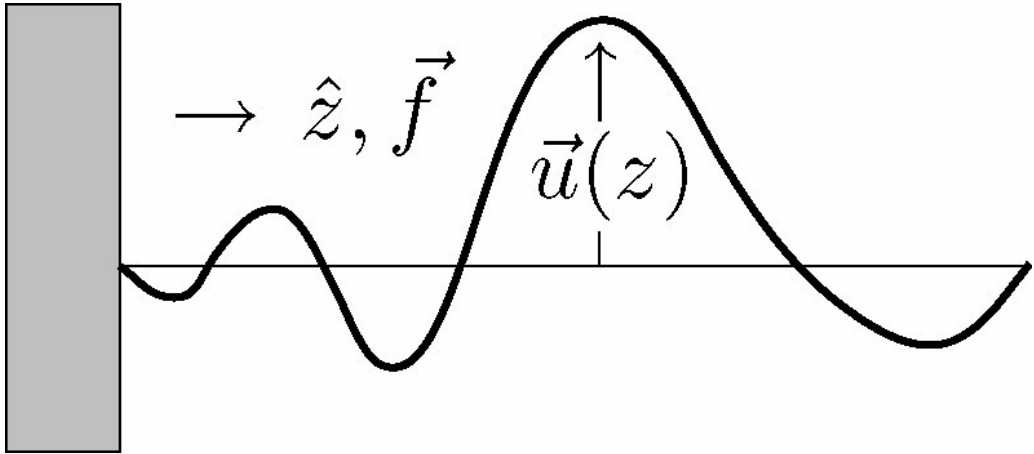

Fig. 4:

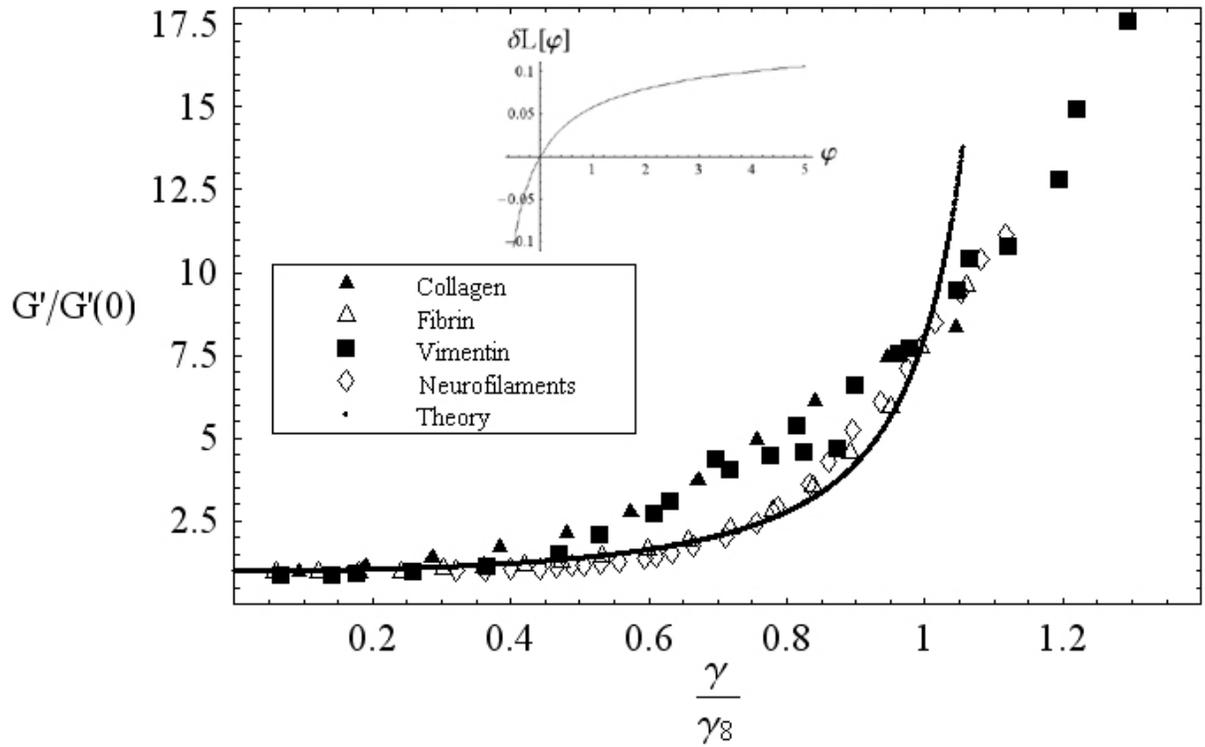

Fig. 5:

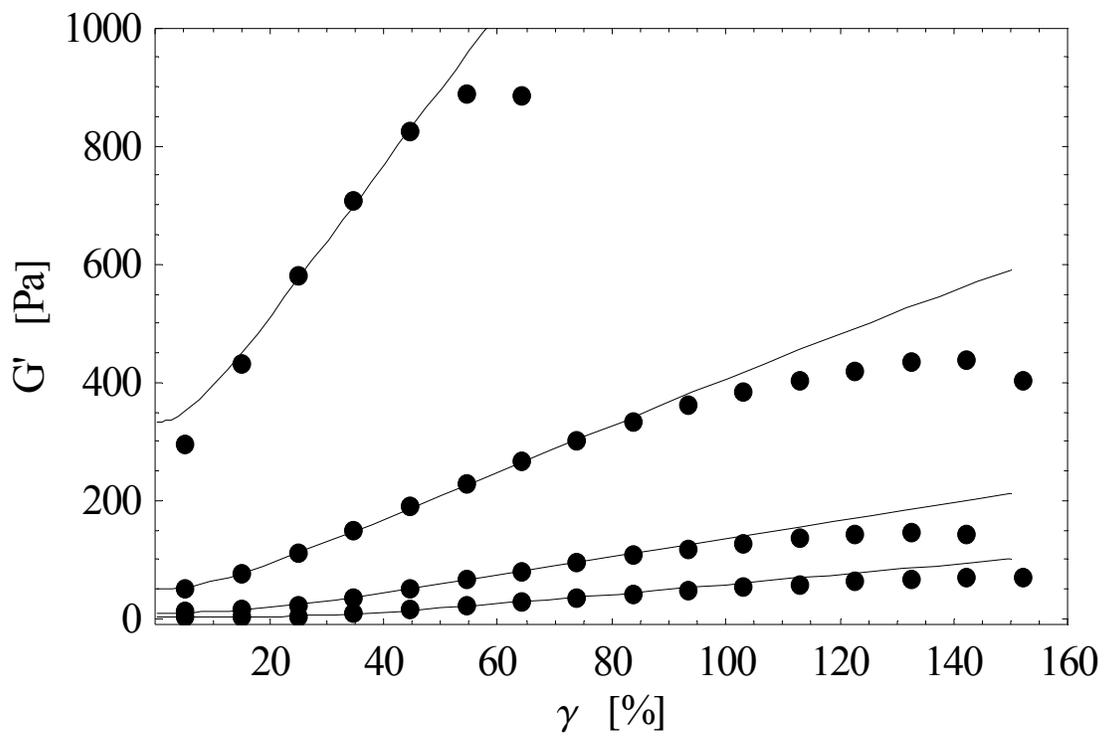